\documentclass[aps,prl,floatfix,twocolumn,showpacs,10pt]{revtex4-1}

\usepackage{graphicx}
\usepackage{dcolumn}
\usepackage{bm}
\usepackage{amssymb}
\usepackage{rotating} 

\begin{document}

\title{Quantum Solver of Contracted Eigenvalue Equations for Scalable Molecular Simulations on Quantum Computing Devices}

\author{Scott E. Smart and David A. Mazziotti}

\email{damazz@uchicago.edu}
\affiliation{Department of Chemistry and The James Franck Institute, The University of Chicago, Chicago, IL 60637}%

\date{Submitted April 21, 2020}


\begin{abstract}

The accurate computation of ground and excited states of many-fermion quantum systems is one of the most consequential, contemporary challenges in the physical and computational sciences whose solution stands to benefit significantly from the advent of quantum computing devices.  Existing methodologies using phase estimation or variational algorithms have potential drawbacks such as deep circuits requiring substantial error correction or non-trivial high-dimensional classical optimization.  Here we introduce a quantum solver of contracted eigenvalue equations, the quantum analogue of classical methods for the energies and reduced density matrices of ground and excited states.  The solver does not require deep circuits or difficult classical optimization and achieves an exponential speed-up of the exact classical algorithms.  We demonstrate the algorithm though computations on both a quantum simulator and two IBM quantum processing units.

\end{abstract}


\maketitle


{\em Introduction:} Quantum computing has the potential to remove the exponential scaling of many-fermion quantum systems by the direct representation and manipulation of quantum states~\cite{McArdle2020, Feynman1982, Lloyd1996, Abrams1997, Abrams1999, Bravyi2002, Aspuru-Guzik2005, Lanyon2010, Du2010, Peruzzo2014, Wang2015, Wecker2015, OMalley2016, McClean2016, Kandala2017, Shen2017, Paesani2017, Lee2018, Kandala2019, Motta2019, Grimsley2019, Wang2019, Ganzhorn2019, McArdle2019, Mazzola2019, Smart2019, Bian2019, Smart2019a, Matsuzawa2020, Wei2020, Ryabinkin2020, Gard2020, Stair2020, Xu2020, Huang2020, Mitarai2020}.  Algorithms for solving the energy eigenvalue equation of many-fermion systems include quantum phase estimation (QPE)~\cite{Paesani2017, Aspuru-Guzik2005}, adiabatic state preparation (ASP)~\cite{Du2010}, and the variational quantum eigensolver (VQE)~\cite{Wang2019, Kandala2017, McClean2016, Wecker2015, Peruzzo2014}.  QPE requires deep circuits with substantial error correction  and ASP utilizes a slow and long time evolution with the computational costs of both methods quickly outpacing the capabilities of near-term quantum computers.  While VQE has shown practical improvements over QPE and ASP, it suffers from high-dimensional classical optimization over a non-ideal surface, typically relying upon derivative-free optimization~\cite{Larson2019} whose computational cost increases rapidly with system size.  Here we introduce a quantum eigenvalue solver that solves a contraction (or projection) of the eigenvalue equation for efficient, scalable molecular simulations on quantum computers.

We develop a novel quantum eigensolver that optimizes the lowest energy eigenvalue by solving a contracted eigenvalue equation.  The projection of the Schr{\"o}dinger equation onto 2-particle transitions from the wave function is known as the 2-particle contracted Schr{\"o}dinger equation (CSE)~\cite{Mazziotti2007, Coleman2000, Colmenero1993a, Nakatsuji1996, Yasuda1997, Mazziotti1998b, Mazziotti1998a, Mazziotti1999a, Mazziotti2002d, Erdahl2007}.   Here we consider the anti-Hermitian part of the CSE known as the 2-particle anti-Hermitian CSE (ACSE)~\cite{Mazziotti2006, Mazziotti2007a, Mazziotti2007b, Mazziotti2007d, Rothman2009, Gidofalvi2009, Sand2015, Alcoba2011, Mukherjee2001, Yanai2006, Li2016}, which has been used in many-electron quantum theory to solve for the ground- and excited-state energies and properties of strongly correlated atoms and molecules~\cite{Smart2018, Schlimgen2017, Sturm2016, Snyder2011, Snyder2011c, Greenman2011, Snyder2010, Mazziotti2008}.  As shown previously, the solution of the ACSE has a close connection to the variational minimization of the energy with respect to a series of 2-body unitary transformations~\cite{Mazziotti2004c, Mazziotti2006, Mazziotti2007a, Mazziotti2007b}.  The gradient of the energy with respect to the 2-body unitary transformations is the residual of the ACSE, and hence, the gradient with respect to these transformations vanishes if and only if the ACSE is satisfied~\cite{Mazziotti2004c, Mazziotti2006, Mazziotti2007a}.  In the classical algorithms the solution of the ACSE for the 2-particle reduced density matrix (2-RDM) is indeterminant without reconstruction of the 3-RDM~\cite{Mazziotti2006, Mazziotti2007a, Mazziotti2007b, Mazziotti2007d, Mazziotti2007g, Rothman2009, Gidofalvi2009, Sand2015, Mazziotti1998b, Mazziotti1999a}.  In the quantum algorithm, however, we show that through the preparation and measurement of the quantum state, the ACSE can be solved for the 2-RDM without any reconstruction or storage of the 3-RDM.  The algorithm exhibits an exponential speedup relative to the exact classical algorithm.

A quantum contracted-eigenvalue-equation solver for solving the ACSE is applied to several problems on IBM quantum computers and an IBM simulator.  On a quantum computer we solve for the ground-state dissociation of the hydrogen molecule.  Both energies and 2-RDMs are computed.  On a one-qubit IBM device we also solve a one-qubit Hamiltonian to demonstrate the trajectory of the solution of the ACSE in iteratively optimizing the ground-state energy.  Lastly, we compute the ground-state dissociation of the linear H$_{3}$ molecule on a quantum simulator.  While the solution of linear H$_{3}$ by the classical algorithm yields a ground-state energy that is limited by the accuracy of the approximate cumulant reconstruction of the 3-RDM, the quantum-computing algorithm yields a ground-state potential energy curve that can be converged to the exact solution for all computed internuclear distances.


{\em Theory:} We begin with the Schr{\"o}dinger equation for a many-electron quantum system
\begin{equation}
({\hat H} - E) | \Psi \rangle = 0 ,
\end{equation}
with the Hamiltonian operator
\begin{equation}
\label{eq:Hhat}
{\hat H} = \sum_{pqst}{^{2} K^{pq}_{st} {\hat a}^{\dagger}_{p} {\hat a}^{\dagger}_{q} {\hat a}^{ }_{t} {\hat a}^{ }_{s}} ,
\end{equation}
where $^{2} K$ is the two-electron reduced Hamiltonian matrix, containing the one- and two-electron integrals, and the second-quantized operators ${\hat a}^{\dagger}_{p}$ and ${\hat a}_{p}$ create and annihilate an electron in the spin orbital $p$, respectively.   The projection (or contraction) of the Schr{\"o}dinger equation onto the space of two-electron transitions generates the CSE~\cite{Mazziotti2007, Coleman2000, Colmenero1993a, Nakatsuji1996, Yasuda1997, Mazziotti1998b, Mazziotti1998a, Mazziotti1999a, Mazziotti2002d}, and the anti-Hermitian part of the CSE produces the ACSE~\cite{Mazziotti2006, Mazziotti2007a, Mazziotti2007b, Mazziotti2007d, Mazziotti2007g, Rothman2009, Gidofalvi2009, Alcoba2011, Sand2015}
\begin{equation}
\label{eq:ACSE}
\langle \Psi | [{\hat a}^{\dagger}_{i} {\hat a}^{\dagger}_{j} {\hat a}^{ }_{l} {\hat a}^{ }_{k} , {\hat H} ] | \Psi \rangle = 0 .
\end{equation}
The ACSE is important for many-electron quantum systems, especially---as we show below---in quantum computing, because its residual contains the gradient for the optimization of many-electron wave functions.

Consider the variational ansatz for the wave function in which the wave function is iteratively constructed from unitary two-body exponential transformations~\cite{Mazziotti2004c, Mazziotti2006, Mazziotti2007a}
\begin{equation}
\label{eq:Psi}
| \Psi_{n+1} \rangle = e^{\epsilon {\hat A}_{n}} | \Psi_{n} \rangle ,
\end{equation}
where ${\hat A}_{n}$ is an anti-Hermitian two-electron operator
\begin{equation}
\label{eq:Ahat}
{\hat A}_{n} = \sum_{pqst}{ ^{2} A^{pq;st}_{n} {\hat a}^{\dagger}_{p} {\hat a}^{\dagger}_{q} {\hat a}^{ }_{t} {\hat a}^{ }_{s}} .
\end{equation}
The energy at the $(n+1)^{\rm th}$ iteration through order $\epsilon$ is given by
\begin{equation}
E_{n+1} = E_{n} + \epsilon \langle \Psi_{n} | [ {\hat H}, {\hat A_{n}}] | \Psi_{n} \rangle .
\end{equation}
Consequently, the gradient of the energy with respect to $^{2} A_{n}$ is
\begin{equation}
\frac{\partial E_{n}}{\partial (^{2} A^{ij;kl}_{n})} = - \langle \Psi_{n} | [{\hat a}^{\dagger}_{i} {\hat a}^{\dagger}_{j} {\hat a}^{ }_{l} {\hat a}^{ }_{k} , {\hat H} ] | \Psi_{n} \rangle .
\end{equation}
From this equation we observe two important facts~\cite{Mazziotti2004c, Mazziotti2006, Mazziotti2007a}: (1) the residual of the ACSE is the negative of the energy gradient with respect to all two-body unitary transformations parametrized by $^{2} {\hat A}_{n}$ and (2) the residual of the ACSE with respect to $\Psi_{n}$ vanishes if and only if the sequence of wave functions has converged at $n$ to a local minimum of the energy.

The ACSE can be solved to compute the 2-RDM directly without storage of the many-electron wave function.  In the algorithm previously implemented on classical computers~\cite{Mazziotti2006, Mazziotti2007a, Mazziotti2007b, Mazziotti2007d, Mazziotti2007g, Rothman2009, Gidofalvi2009, Alcoba2011, Sand2015}, the wave function at the $n^{\rm th}$ iteration is substituted into the definition of the 2-RDM
\begin{equation}
\label{eq:D2}
^{2} D^{pq}_{st} = \langle \Psi | {\hat a}^{\dagger}_{p} {\hat a}^{\dagger}_{q} {\hat a}^{ }_{t} {\hat a}^{ }_{s} | \Psi \rangle
\end{equation}
to yield an expression for the 2-RDM at the $(n+1)^{\rm th}$ iteration
\begin{equation}
^{2} D^{pq;st}_{n+1} = ^{2} D^{pq;st}_{n} + \epsilon \langle \Psi_{n} | [ {\hat a}^{\dagger}_{p} {\hat a}^{\dagger}_{q} {\hat a}^{ }_{t} {\hat a}^{ }_{s}, {\hat A}_{n}] | \Psi_{n} \rangle .
\end{equation}
where the operator ${\hat A}_{n}$ can be selected to be the residual of the ACSE, which causes the 2-RDM to follow the energy's gradient towards its minimum
\begin{equation}
^{2} A^{ij;kl}_{n} = \langle \Psi_{n} | [{\hat a}^{\dagger}_{i} {\hat a}^{\dagger}_{j} {\hat a}^{ }_{l} {\hat a}^{ }_{k} , {\hat H} ] | \Psi_{n} \rangle  .
\end{equation}
By using the fact that ${\hat A}_{n}$ and ${\hat H}$ are two-body operators in Eqs. (\ref{eq:Ahat}) and (\ref{eq:Hhat}), the 2-RDM at the $(n+1)^{\rm th}$ iteration can be expressed as a linear functional of the 1-, 2-, and 3-RDMs at the $n^{\rm th}$ iteration.  The indeterminacy in these recursion relations for the 2-RDM can be removed by reconstructing the 3-RDM approximately from the 2-RDM~\cite{Mazziotti2006, Mazziotti2007a, Mazziotti2007b, Mazziotti2007d}.  For example, the cumulant part of the 3-RDM in its cumulant expansion can be neglected or approximated to provide a reconstruction of the 3-RDM in terms of the 2-RDM~\cite{Mazziotti2006, Mazziotti1998a, Mukherjee2001}.

We propose a novel algorithm for solving the ACSE for the 2-RDM on the quantum computer, which is shown in Table~I. While the classical computer uses matrices and vectors to represent quantum states, the quantum computer allows us to prepare a form of the quantum state itself in terms of qubits where the scaling of the preparation is non-exponential~\cite{Nielsen2010}.  Utilizing this capability, we prepare the wave function at the $(n+1)^{\rm th}$ in Eq.~(\ref{eq:Psi}) on the quantum computer (Step 3 of Table~I) and perform measurements of its 2-RDM's matrix elements in Eq.~(\ref{eq:D2}) on the quantum computer (Step 4).  In Step~5 we optimize the parameter $\epsilon$ by minimizing the energy by a model-trust Newton's method~\cite{Nocedal2006}.  Before we can perform the preparation in Step~3, however, we need to compute the $^{2} A$ matrix by evaluating residual of the ACSE.  While we could evaluate the ACSE on the classical computer using Eq.~(\ref{eq:ACSE}) with cumulant reconstruction of the 3-RDM, we can compute the residual of the ACSE on the quantum computer directly without a formal approximation.   We prepare the auxiliary state $|\Lambda_{n}\rangle$ in Step~1 of the algorithm
\begin{equation}
| \Lambda_{n} \rangle = e^{i \delta {\hat H}} | \Psi_{n} \rangle
\end{equation}
where $\delta$ is a small nonnegative parameter and measure the imaginary part of its 2-RDM on the quantum computer which gives us the residual of the ACSE---the elements of the $^{2} A$ matrix
\begin{equation}
^{2} A^{ij;kl}_{n} = {\rm Im}\langle \Lambda_{n} | {\hat a}^{\dagger}_{i} {\hat a}^{\dagger}_{j} {\hat a}^{ }_{l} {\hat a}^{ }_{k}  | \Lambda_{n} \rangle .
\end{equation}
Steps (1-4) can be repeated until convergence.  Because we are following the gradient, for a suitably small choice of $\epsilon$ the algorithm is guaranteed to converge.  The algorithm can be initiated with any initial wave function including the mean-field (Hartree-Fock) wave function.  The only errors in the quantum solution of the ACSE for the 2-RDM arise from noise on the quantum computer, and hence, even with an initial Hartree-Fock wave function, the algorithm can treat strongly correlated molecular quantum systems.


\begin{table*}[t!]
  \caption{\normalsize Quantum ACSE algorithm for 2-RDM optimization.}
  \label{t:QACSE}
  \begin{ruledtabular}
  \begin{tabular}{l}
  {\bf Algorithm: Quantum ACSE method for 2-RDM optimization} \\
  \hspace{0.5in} {\em Given} $n=0$ {\em and} $0 < \delta \le 1$. \\
  \hspace{0.5in} {\em Choose initial wave function} $| \Psi_{0} \rangle$. \\
  \hspace{0.5in} {\em Repeat until} $||^{2} A_{n}||$ {\em is small.} \\
  \hspace{1.0in} {\em {\bf Step 1:} Prepare} $| \Lambda_{n} \rangle$ {\em from} $| \Lambda_{n} \rangle = e^{i \delta {\hat H}} | \Psi_{n} \rangle$, \\
  \hspace{1.0in} {\em {\bf Step 2:} Measure} $^{2} A_{n}$ {\em from} $^{2} A^{ij;kl}_{n} = {\rm Im}\langle \Lambda_{n} | {\hat a}^{\dagger}_{i} {\hat a}^{\dagger}_{j} {\hat a}^{ }_{l} {\hat a}^{ }_{k}  | \Lambda_{n} \rangle$, \\
  \hspace{1.0in} {\em {\bf Step 3:} Prepare} $| \Psi_{n+1} \rangle$ {\em from} $ | \Psi_{n} \rangle = e^{ \epsilon {\hat A}} | \Psi_{n} \rangle$, \\
  \hspace{1.0in} {\em {\bf Step 4:} Measure} $^{2} D_{n+1}$ {\em from} $^{2} D^{pq;st}_{n+1} = \langle \Psi_{n+1} | {\hat a}^{\dagger}_{p} {\hat a}^{\dagger}_{q} {\hat a}^{ }_{t} {\hat a}^{ }_{s} | \Psi_{n+1} \rangle$, \\
   \hspace{1.0in} {\em {\bf Step 5:} Iterate Steps 3 and 4 to minimize the energy with respect to $\epsilon$},\\
  \hspace{1.0in} {\em {\bf Step 6:} Set} $n = n+1$.
  \end{tabular}
  \end{ruledtabular}
\end{table*}


{\em Results:} To illustrate the solution of the ACSE on the quantum computer, we apply the quantum ACSE algorithm to three applications: the solution of a generic one-qubit Hamiltonian and the dissociation of the H$_{2}$  and H$_{3}$ molecules.  Solutions of the one-qubit and H$_{2}$ Hamiltonians  are performed on the one- and five-qubit IBM quantum computers Armonk and Ourense, respectively~\footnote{Supplemental Material contains additional details of the computations and specifications of the quantum computers}.  The dissociation of H$_{3}$ is implemented on a quantum simulator to probe the method's accuracy in the absence of noise.


\begin{figure}
\begin{center}
\includegraphics[scale=0.35]{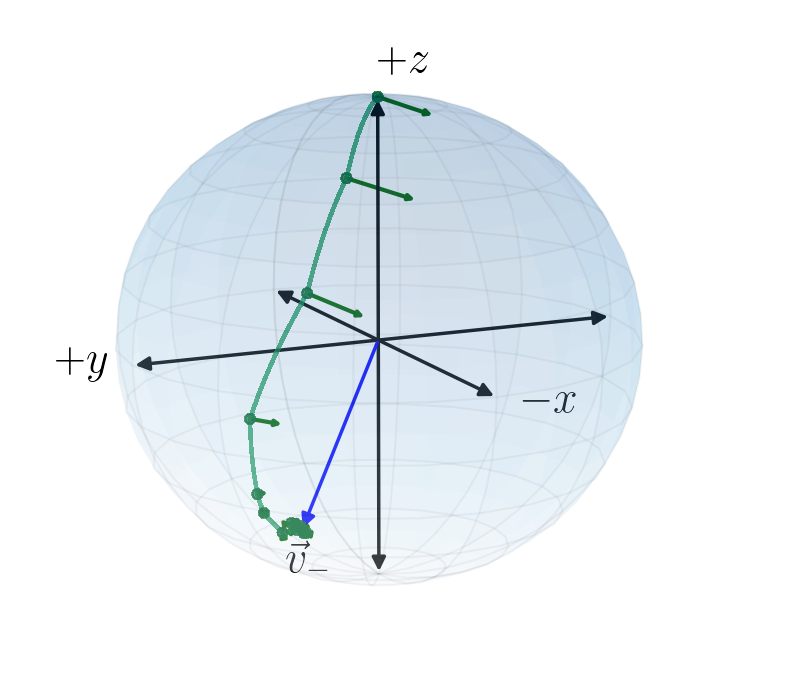}
\end{center}
\caption{For a 1-qubit Hamiltonian the solution of the quantum ACSE converges to the ground state, indicated by $v_{-}$, in about 8 iterations on a 1-qubit IBM quantum computer.}
\label{fig:H1}
\end{figure}

We first examine the solution of a one-qubit Hamiltonian ${\hat H} = \frac{1}{2} \left ( {\hat \sigma}_{x} - {\hat \sigma}_{y} + {\hat \sigma}_{z} \right )$ where ${\hat \sigma}_{x}$, ${\hat \sigma}_{y}$, and ${\hat \sigma}_{z}$ are the Pauli matrices in the $x$, $y$, and $z$ directions.  In the basis of the Bloch sphere we can express the Hamiltonian and the initial density matrix as vectors $(1,-1,1)$ and $(0,0,1)$, respectively.  The minimization of the ground-state energy by the quantum ACSE algorithm is .  Beginning at the initial density-matrix vector, the solution from the ACSE reaches the ground-state vector, indicated by $v_{-}$, in approximately 8 iterations on the quantum computer, as shown in Fig.~1.  The quantum solution of the ACSE provides an efficient mechanism for moving along the surface of the Bloch sphere, where the pure states exist, to reach the ground-state solution. In the special case of the Bloch sphere the ${\hat A}$ operator can be computed as a Bloch vector from the cross product of the Hamiltonian vector with the density-matrix vector; geometrically, from the definition of the cross product the ${\hat A}$ vector, indicated in Fig.~1 at each iteration by an arrow, is orthogonal to the plane formed by the Hamiltonian and density-matrix vectors.

\begin{figure}
\begin{center}
\includegraphics[scale=0.5]{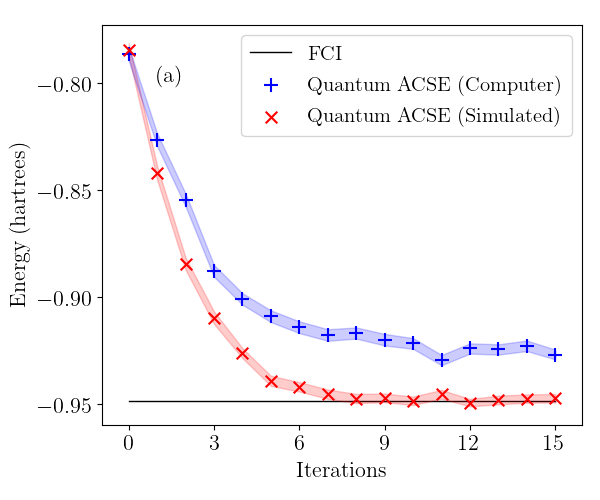}
\includegraphics[scale=0.5]{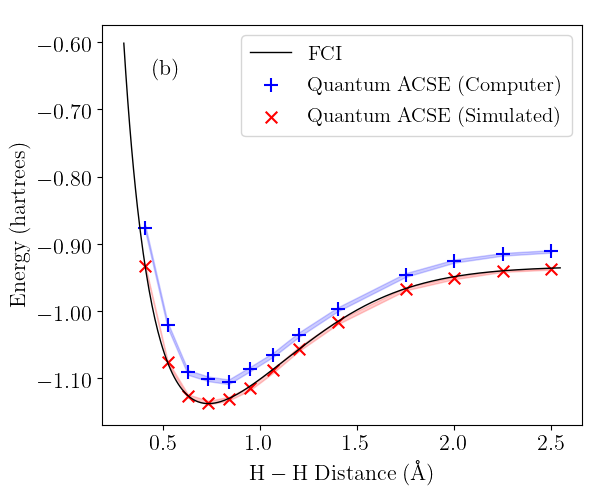}
\end{center}
\caption{For the H$_{2}$ molecule the figure shows (a) the energy at each iteration in the solution of the ACSE at an internuclear distance $R$ of 2~\AA\ and (b) the energy dissociation curve of the molecule.  The error in the ACSE on the quantum computer, due to noise, is fairly uniform throughout the dissociation, indicating that the ACSE captures the spin entanglement.}
\label{fig:H2}
\end{figure}

Second, we compute the dissociation of the hydrogen molecule in a minimal Slater-type orbital (STO-3G) basis set.  On the quantum computer the molecule is represented in the ACSE algorithm by a two-qubit compact mapping.  The energy at each iteration in the solution of the ACSE for H$_{2}$ at 2~\AA\ is shown in Fig.~2a.  The ACSE energy from a quantum simulator converges to the energy from full configuration interaction (FCI) in about 9-to-10 iterations.  The ACSE energy on the quantum computer converges in approximately the same number of iterations to an energy that is approximately 25~mhartrees higher than the FCI energy.  This error is due to the noise present on the quantum computer; in fact, a nearly identical curve in the iterations is generated by a quantum simulator with the QISKIT noise model, a noise model based on the device T1, T2, and readout parameters.  Figure~2b shows the energy dissociation curve of the hydrogen molecule.  While the noise error is visible in the potential energy curve from the ACSE, the error is importantly uniform throughout the curve, indicating that the ACSE algorithm is capturing the significant electron correlation from spin entanglement in the dissociation region.

\begin{figure}
\begin{center}
\includegraphics[scale=0.5]{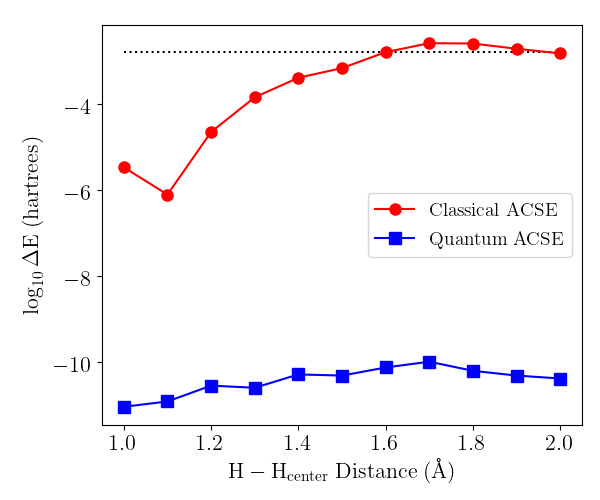}
\end{center}
\caption{The error in the potential energy curves from the equal-bond dissociation of the H$_{3}$ molecule are shown for the classical and quantum ACSE algorithms with the quantum ACSE being more accurate by 6 orders of magnitude.  Dotted line indicates ``chemical accuracy'' (1~kcal/mol).}
\label{fig:H3}
\end{figure}

Finally, we calculate the dissociation of the H$_{3}$ molecule in the minimal Slater-type orbital (STO-3G) basis set on a quantum simulator without noise.  The purpose of this calculation is to examine the accuracy of the quantum ACSE algorithm on an ideal, noise-free quantum computer.  Stretching the two bonds of the molecule equally causes a Mott metal-to-insulator transition with the stretched geometry being highly correlated due to nontrivial spin entanglement~\cite{Smart2019a}.  The energy errors from the classical and quantum ACSE algorithms, relative to the FCI energy, are shown in Fig.~3.  Most strikingly, the energies from the quantum ACSE are about six orders of magnitude more accurate than the energies from the classical ACSE.  While the classical ACSE algorithm requires an approximate cumulant-based reconstruction~\cite{Mazziotti1998a} of the 3-RDM from the 2-RDM, the quantum ACSE algorithm does not require any reconstruction approximation because the updates of the $^{2} S$ and $^{2} D$ matrices are performed, in principle exactly, through a combination of state preparation and tomography.  Figure~3 also shows that the quantum solution of the ACSE remains accurate at stretched bond distances where the electron correlation--- the deviation from the Hartree-Fock solution---is significant.


{\em Discussion and Conclusions:} Key features of the quantum algorithm for solving the ACSE include: (1) computation of the energy and 2-RDM without any approximate reconstruction of higher RDMs as in the classical algorithm and (2) evaluation of the energy gradient--- residual of the ACSE---on the quantum computer for accurate and efficient gradient-based optimization. The ACSE algorithm's computation of the gradient on the quantum computer offers a significant advantage over VQE algorithms~\cite{Wang2019, Kandala2017, McClean2016, Wecker2015, Peruzzo2014} that approximate the gradient on the classical computer by derivative-free optimization methods~\cite{Larson2019} like the simplex method that are limited to hundreds of degrees of freedom.  The quantum ACSE algorithm is also much faster to converge than adiabatic algorithms like ASP~\cite{Du2010} due to its gradient-based optimization.  Finally, it has much lower tomography costs than other methods for accelerating ASP like Lanczos-based imaginary-time evolution methods~\cite{McArdle2019,Motta2019}.  Unlike unitary coupled cluster which uses a single unitary exponential transformation of commuting operators~\cite{Shen2017}, the ACSE method represents the wave function as a product of unitary exponential transformations of two-body, non-commuting operators that represent the higher excitations as non-trivial products of two-body operators; furthermore, the ACSE's iterative approximation to the construction of the wave function decreases Trotterization errors, errors from the application of Trotter's formula for representing the exponential transformation on the quantum computer.

The quantum algorithm for solving the ACSE provides a direct computation of ground-state energies and 2-RDMs with an efficient generation of the search direction from the ACSE residual. In the context of quantum algorithms, it has the benefits of good ansatz depth, modest tomography requirements, and no derivative-free classical optimization.  Importantly, while the focus here is on the solution of many-fermion systems, the ACSE algorithm is also applicable to solving many-boson systems as well as many-qubit systems governed by arbitrary $p$-body interactions.  Future work will also explore the application of the ACSE algorithm to electronic excited states and active-space calculations for the treatment of strong electron correlation in larger molecules.  Because the quantum ACSE algorithm is an iterative approach to computing the $N$-representable 2-RDM~\cite{Mazziotti2012b,Schilling2013,Piris2017a} of a given eigenstate, it offers a polynomial-scaling approach to computing energies and properties of strongly correlated many-fermion quantum systems on both near-to-intermediate-term and future quantum devices with applications across quantum chemistry and physics.

\begin{acknowledgments}

D.A.M. gratefully acknowledges the Department of Energy, Office of Basic Energy Sciences Grant DE-SC0019215, the U.S. National Science Foundation Grant CHE-1565638, the U.S. Army Research Office (ARO) Grant W911NF-16-1-0152, and IBM Q.  The views expressed are of the authors and do not reflect the official policy or position of IBM or the IBM Q team.

\end{acknowledgments}

\bibliography{QComputers,CSE,tACSE,aACSE,oACSE,Cumulant,Opt}


\end{document}